\documentclass{article}

 \usepackage[nonatbib, preprint]{neurips_2021}

\usepackage[utf8]{inputenc} 
\usepackage[T1]{fontenc}    
\usepackage{hyperref}       
\usepackage{url}            
\usepackage{booktabs}       
\usepackage{nicefrac}       
\usepackage{microtype}      
\usepackage{xcolor}         
\usepackage{graphicx,caption,amsmath,amssymb,amsfonts,lipsum,cite,soul,url, subcaption}
\usepackage[ruled,titlenumbered]{algorithm2e}

\newcommand{\real}{\mathbb{R}}

\newcommand{\vo}[1]{{\boldsymbol{#1}}}

\newcommand{\x}{\vo{x}}

\newcommand{\Th}{\vo{\theta}}
\renewcommand{\u}{\vo{u}}
\newcommand{\w}{\vo{w}}

\newcommand{\xdot}{\dot{\x}}
\newcommand{\divergence}[1]{\vo{\nabla}.\left(#1\right)}
\newcommand{\eqnlabel}[1]{\label{eqn:#1}}

\title{Optimal Transport Based Refinement of Physics-Informed Neural Networks}

\author{%
  Vaishnav Tadiparthi 
  \\
  Department of Aerospace Engineering\\
  Texas A\&M University\\
  College Station, TX 77843 \\
  \texttt{vaishnavtv@tamu.edu} \\
  \And
  Raktim Bhattacharya \\
  Department of Aerospace Engineering\\
  Texas A\&M University\\
  College Station, TX 77843 \\
  \texttt{raktim@tamu.edu} \\
}

\begin{document}

\maketitle

\begin{abstract}
  In this paper, we propose a refinement strategy to the well-known Physics-Informed Neural Networks (PINNs) for solving partial differential equations (PDEs) based on the concept of Optimal Transport (OT).
  Conventional black-box PINNs solvers have been found to suffer from a host of issues: spectral bias in fully-connected architectures, unstable gradient pathologies, as well as difficulties with convergence and accuracy.
  Current network training strategies are agnostic to dimension sizes and rely on the availability of powerful  computing resources to optimize through a large number of collocation points.
  This is particularly challenging when studying stochastic dynamical systems with the Fokker-Planck-Kolmogorov Equation (FPKE), a second-order PDE which is typically solved in high-dimensional state space.
  While we focus exclusively on the stationary form of the FPKE, positivity and normalization constraints on its solution make it all the more unfavorable to solve directly using standard PINNs approaches.
  To mitigate the above challenges, we present a novel training strategy for solving the FPKE using OT-based sampling to supplement the existing PINNs framework.
  It is an iterative approach that induces a network trained on a small dataset to add samples to its training dataset from regions where it nominally makes the most error.
  The new samples are found by solving a linear programming problem at every iteration.
  The paper is complemented by an experimental evaluation of the proposed method showing its applicability on a variety of stochastic systems with nonlinear dynamics.

\end{abstract}

\section{Introduction}
The problem of uncertainty propagation through nonlinear dynamical systems with stochastic excitations or uncertain parameters arises in a variety of fields, ranging from statistical physics to economy and finance \cite{escobedo1999radiation, furioli2017fokker}.
The exact description of the time evolution of the probability density function (pdf) $\rho(t,\x)$ corresponding to the state $\x$, can be analysed by studying the Fokker-Planck-Kolmogorov equation (FPKE) \cite{fuller1969analysis, risken1996fokker, elgin1984fokker}.
It is a partial differential equation (PDE) whose analytical solutions for different dynamical systems are difficult to obtain except under very special circumstances, for eg., systems with linear dynamics and a small class of nonlinear low-dimensional systems \cite{caughey1982steady, caughey1967response}. In this paper, we are concerned with solving the steady-state form of the FPKE for any nonlinear system.

There are several numerical techniques to solve the FPKE like the finite element method \cite{spencer1993numerical, naprstek2014finite, galan2007stochastic}, finite difference method \cite{jiang2015new, sepehrian2015numerical},
path integral methods \cite{drozdov1998accurate, xu2019path}, that have emerged from the literature on solving PDEs.
While these methods perform reliably for systems with low dimensions , they do not scale well and are known to become computationally intractable for high-dimensional systems.
Refinement techniques \cite{kumar2009partition, kumar2006multi} to combat the curse of dimensionality have been introduced in the past.
The orthodox way to study uncertainty propagation is by using the Monte Carlo method \cite{johnson1997observations, harnpornchai1999stochastic}: to generate a large number of sample paths from a family of test points, propagate the nonlinear system forward exactly, and study the statistics at the time-steps of choice.
While this certainly provides a reliable solution to the governing PDE, it too depends heavily on the amount of computing resources available to generate large amounts of data, plagued again by the curse of dimensionality.
Alternative methods to analyse uncertainty propagation involve linearization of the nonlinear system in consideration \cite{pradlwarter2001non}, and the choice to ignore the higher-order terms can lead to significant errors.

This is where the universal approximation power of neural networks is coming to the fore.
With novel developments in architecture design and training strategies, they have been tremendously successful in the data-rich domains of computer vision, speech processing and the like, revolutionizing the way we build predictive models of complex systems today.
In the function approximation world of physics-informed scientific computing, there has been remarkable progress in the way these black-box models are being harnessed to solve solve problems by endowing them with prior domain knowledge and appropriate inductive biases \cite{sun2020surrogate, raissi2019deep, fang2019deep}.

Of late, the dominant school of thought to train neural networks to solve differential equations is by introducing penalty terms into the loss functions that constrain the obtained solutions to satisfy some sets of governing conditions \cite{raissi2018deep}.
Specifically, physics-informed neural networks (PINNs) solve PDEs by enforcing a composition of constraints coming from the PDE and associated initial and boundary conditions.

Unfortunately however, PINNs have been found to suffer from a host of issues: spectral bias from fully-connected architectures, instabilities during training, convergence and accuracy issues with "stiff" PDEs \cite{wang2020understanding, mcclenny2020self, wang2020and}. Numerous modifications to the baseline PINNs algorithm have been proposed and most of them revolve around adaptation mechanisms to strategically weight the different loss functions in the composite loss to account for differing gradients used during training.

Using the baseline PINNs for solving the FPKE is also not as straightforward. The FPKE is a formidable challenge to solve using a generic neural network because the resulting pdf has to satisfy two unique properties: positivity over the entire domain and a normalization condition.
While positivity can be ensured through an appropriate choice of architecture, enforcing normalization through a penalty function \cite{xu2020solving} introduces another gradient pathology that can be hard to resolve.

To overcome these issues, we transform the stationary FPKE to an equivalent PDE that allows us to use the conventional PINNs framework.
A state-independent diffusion vector allows us to generalize the solution to the true FPKE as the exponential of a potential function \cite{risken1996fokker}, i.e., $\rho(\x) = N_0 \textrm{exp}({-\eta(\x)})$, where $\eta(\x)$ is the potential and  $N_0$ is a normalizing coefficient.
This transformation implicitly preserves positivity, and normalization can be performed using numerical integration over the entire domain after the PDE in $\eta(\x)$ has been solved.

Further, to overcome any instability issues that may arise during training, we propose a refinement strategy based on the theory of optimal transport (OT) \cite{villani2008optimal} to actively choose collocation points that improve the training data regime over which the solution is learnt.
The OT framework allows the synthesis of optimal mappings between functions on some measure spaces with respect to a variety of cost functions, depending on the nature of the problem.
The theory links very well with stochastic dynamical systems, and has been used with great success in the past for solving Bayesian filtering problems \cite{das2020optimal}.

Typically, collocation points are chosen through sampling from a pre-determined distribution or placed uniformly across the grid in consideration.
Uniform grid distributions do not scale well with increasing dimensionality, nor do they account for the efficacy of the obtained solution over the entire domain.
Alternatively, we seek to generate these points by examining the underlying distribution of the error regime from a nominally trained network.
We solve a linear programming problem based on the theory of optimal transport to obtain these new points at the regions where the network fails the most, and iteratively train the network to a reasonable degree of accuracy.
This can work in tandem with the adaptive weight tuning strategies proposed in the last few years, remarkably improving the utility of a small training set and reducing the computational overhead of training a highly complex network over each and every point in a very fine mesh.

The primary contributions of this paper are:
\begin{itemize}
  \item A novel iterative training scheme based on the theory of Optimal Transport to generate collocation points for PINNs in regions of high error, thereby improving the sample efficiency of a conventional PINNs solver

  \item A PINNs framework to solve the stationary Fokker-Planck equation for any dynamical system through a convenient transformation, thus ensuring the requisite conditions of the obtained probability density function
\end{itemize}

The paper is organized as follows: in section 2, we briefly describe the original PINNs formulation and the Fokker-Planck-Kolmogorov equation for any stochastic dynamical system.
Section \ref{sec_3.1} details the reduced formulation for the FPKE that allows the conventional PINNs framework to yield satisfactory results.
In section \ref{sec_ot}, we elaborate upon the proposed sampling strategy using Optimal Transport.
This is succeeded by numerical simulations on a few systems, demonstrating the prowess of the formulations proposed.
A helpful algorithm has been provided for reference.
The paper ends with a summary of the findings and some questions posed for further research.

\section{Background}

\subsection{Physics Informed Neural Networks}
Consider any nonlinear PDE of the form:
\begin{subequations}
  \begin{align}
    &\mathcal{N}_{t,\x} [\u(t,\x)] = 0, \x \in \Omega, t \in [0,T], \\
    &  \u(t, \x) = g(t, \x), \x \in  \partial \Omega, t \in [0,T],\\
    &  \u(0, \x) = h(\x), \x \in \Omega.
  \end{align}
\eqnlabel{pde_pinn}
\end{subequations}
where $\mathcal{N}_{t,\x}$ is a spatio-temporal differential operator.
$\x \in \Omega$ is a $n$-dimensional spatial vector-variable operating in a domain $\Omega \in \real^n$ and $t$ is time.
$\partial \Omega$ refers to the boundary of the domain $\Omega$.
Following the work of \cite{raissi2019deep}, the objective is to approximate $\u(t,\x)$ by a fully-connected neural network $\u_{\Th}(t,\x)$ with inputs $t$ and $\x$, parameterized by $\Th$.
Define an additional residual network $r_{\Th}(t,\x)$ with the same weights $\Th$ such that:
\begin{align*}
  r_{\Th}(t,\x) = \mathcal{N}_{t,\x} [\u_{\Th}(t,\x)].
\end{align*}

The optimal set of weights is obtained by training the neural network on a composite loss function of the form:
\begin{align}
  \mathcal{L}(\Th) =   \mathcal{L}_{r}(\Th) +   \mathcal{L}_{b}(\Th) + \mathcal{L}_{i}(\Th),
\end{align}
where
\begin{subequations}
  \begin{align}
    \mathcal{L}_{r}(\Th) &= \frac{1}{N_r}\sum_{j=1}^{N_r} |r_{\Th}(t,\x_r^j)|^2 \eqnlabel{eqn_pdeError},\\
    \mathcal{L}_{b}(\Th) &= \frac{1}{N_b}\sum_{j=1}^{N_b} |\u_{\Th}(t,\x_b^j) - g(t, \x_b^j)|^2, \\
    \mathcal{L}_{i}(\Th) &= \frac{1}{N_r}\sum_{j=1}^{N_r} |\u_{\Th}(\x_i^i) - h(\x_i^j)|^2.
  \end{align}
  \eqnlabel{loss_pinn}
\end{subequations}
Here, $\mathcal{L}_r(\Th)$ corresponds to the loss function on the residual from the original PDE, and $\mathcal{L}_b(\Th)$ and $\mathcal{L}_i(\Th)$ relate to losses on the boundary and initial conditions respectively.


\subsection{Fokker-Planck Equation}

The probability density function of stochastic differential equations is governed by the Fokker-Planck equation.
The stationary response of nonlinear dynamical systems can be obtained by solving the steady-state Fokker-Planck equation.
It is a linear partial differential equation (PDE) whose analytical solutions have been derived for a restricted class of systems.

Assuming an $n$-dimensional stochastic differential equation (SDE) of the form
$$
\xdot = \vo{F}(\x) + \Lambda (\x, t) \Gamma,
$$
with $\x \in \real^n$, $\Lambda$ is an $n \times m$ matrix and $\Gamma$ is a vector of uncorrelated standard Gaussian white-noise processes.
We are interested in the evolution of the state PDF (probability density function)  $\rho(t,\x)$ given by the Fokker-Planck equation,
\begin{align}
{\partial \rho (t,\x) \over \partial t} + \divergence{\vo{F}(\x)\rho(t,\x)} -
 \sum_{i=1}^n \sum_{j=1}^{n} \frac{\partial^2}{\partial x_i \partial x_j} (D_{ij}(\x, t) \rho(t,\x))= 0. \nonumber\\
 {\partial \rho (t,\x) \over \partial t} + \sum_{i=1}^n \frac{\partial}{\partial x_i} (\vo{F}(\x) \rho(t, \x)) -
  \sum_{i=1}^n \sum_{j=1}^{n} \frac{\partial^2}{\partial x_i \partial x_j} (D_{ij}(t,\x) \rho(t,\x))= 0.
 \eqnlabel{fpke}
\end{align}
where $D = \frac{1}{2} \Lambda \Lambda^T$.

The probability density function $\rho(t,\x)$ gives the probability of being in a differential element $(\x, \x + d\x)$ of the phase plane at time $t$.
It is subject to the normalization condition:
\begin{align}
	\int_{\x} \rho(t, \x) d\x= 1.
	\eqnlabel{normCond}
\end{align}

The boundary conditions are given by $\rho \rightarrow 0$ as $|x_i| \rightarrow \infty$ $\forall i = 1,2,\dots,n$.
This is inferred from the normalization condition, but alternative conditions like zero flux at infinity can be imposed as well \cite{langtangen1991general}.

Further, assuming a constant $\Lambda$ reduces \eqref{eqn:fpke} to:
\begin{align}
	{\partial \rho (t,\x) \over \partial t} + \sum_{i=1}^n \frac{\partial}{\partial x_i} (\vo{F}(\x) \rho(t, \x)) -
   \sum_{i=1}^n \sum_{j=1}^{n} D_{ij}(t,\x) \frac{\partial^2 \rho(t,\x)}{\partial x_i \partial x_j} = 0.
\end{align}

In the steady state, the equation is:
\begin{align}
	\sum_{i=1}^n \frac{\partial}{\partial x_i} (\vo{F}(\x) \rho(\x)) -
   \sum_{i=1}^n \sum_{j=1}^{n} D_{ij}(\x) \frac{\partial^2 \rho(\x)}{\partial x_i \partial x_j}= 0.
	 \eqnlabel{fpke_ss}
\end{align}

We seek to find the solution $\rho_{\Th}(\x)$ to \eqref{eqn:fpke_ss} using PINNs.
Alternatively, \eqref{eqn:fpke_ss} can be expressed as
\begin{align}
\mathcal{N}_{\x}[\rho_{\Th}(\x)] = 0. \eqnlabel{fpke_ss_op}
\end{align}
It is a linear second-order parabolic PDE for which, a typical PINNs solver will return a trivial zero solution for $\rho(\x)$ during training.
We circumvent this problem by transforming the PDE to an equivalent PDE after a change of variables.
The resulting PDE can be solved in the conventional PINNs framework.
Since we are solving the steady-state FPKE, the composite loss function is simply,
\begin{align}
  \mathcal{L}(\Th) =   \mathcal{L}_{r}(\Th) +   \mathcal{L}_{b}(\Th).
\end{align}

\section{Proposed Formulation}
\subsection{Steady-state FPKE in $\eta$}
\label{sec_3.1}

From \cite{risken1996fokker}, the stationary solution to the standard FPKE can be expressed as:
\begin{align}
\lim_{t \rightarrow \infty} \rho(\x) = N_0 \, \textrm{exp}({-\eta(\x)}),
\end{align}
where $\eta(\x)$ is the potential function associated with the stationary solution to the Fokker-Planck equation.
$N_0$ is a normalizing constant that ensures: $\lim_{t \rightarrow \infty} \int_{\x} \rho(\x) d\x = 1$.

This implies,
\begin{align*}
	\frac{\partial \rho}{\partial x_i} = -\textrm{exp}({-\eta(\x)}) \frac{\partial \eta(\x)}{\partial x_i}, \text{ and }
	\frac{\partial^2 \rho}{\partial x_i \partial x_j} =
	-\textrm{exp}({-\eta(\x)}) \frac{\partial^2 \eta(\x)}{\partial x_i \partial x_j} + \textrm{exp}({-\eta(\x)}) \frac{\partial \eta(\x)}{\partial x_i} \frac{\partial \eta(\x)}{\partial x_j}.
\end{align*}

Consequently, \eqref{eqn:fpke_ss} transforms to

%
\begin{align}
	 \sum_{i=1}^n \left(\frac{\partial F_i(\x)}{\partial x_i}  - F_i(\x) \frac{\partial \eta(\x)}{\partial x_i} \right)   - \sum_{i=1}^n \sum_{j=1}^{n}
	\left[ D_{ij} \left(  \frac{\partial^2 \eta(\x)}{\partial x_i \partial x_j} +  \frac{\partial \eta(\x)}{\partial x_i}  \frac{\partial \eta(\x)}{\partial x_j} \right)
	  \right]
	= 0.
\eqnlabel{fkp_pde}
\end{align}

The solution to this PDE can now be obtained by using conventional PINNs methods.

\subsection{Obtaining Collocation Points using Optimal Transport}
\label{sec_ot}
The baseline PINNs solver has several modes of failure: instability during training \cite{mcclenny2020self, wang2020and}, stiffness in gradient flow dynamics arising from imbalance in loss gradient magnitudes \cite{wang2020understanding}.
Most recent work in the realm of PINNs has been to mitigate issues of this sort by introducing modifiable weights to these loss functions, thus altering the loss function during training to arrive at a reasonably good approximation.

Typically, collocation points used to train the network are generated either on a uniform grid or sampled from a chosen distribution.
The distributions are pre-determined and do not account for the efficacy of the obtained solution across the entire domain.
We propose an iterative formulation in which we use the theory of Optimal Transport (OT) to add points to the training dataset in regions where the nominal PINNs solution fails the most.

All OT problems involve two probability measures $\mu$ and $\nu$ defined on some space $\Omega$ and a non-negative cost function $c(.)$ on $\Omega \times \Omega$.
The choices of $\Omega$ and $c(.)$ subjectively depend on the problem being considered.
Let $\mathcal{D}_{\mu}$ denote the support of the measure $\mu$.
The objective is to find a map $\phi: \mathcal{D}_{\mu} \rightarrow \mathcal{D}_{\nu}$ such that the total cost transporting a single element $\x$ to $\phi(\x)$ is minimized.
This cost is given by: $\inf_{\phi(.)} \int c(\x, \phi(\x)) \mu(d\x)$.
In the context of our proposed formulation, $\mu$ is a uniform distribution over the domain, whereas $\nu$ denotes the equation error distribution computed from the nominal solution at each step.

This transport map can be approximated by a linear map $\vo{\Phi}$, represented as a matrix $\vo{\Phi} := [\phi_{ij}]$ such that:
\begin{align}
  \x_{j,k}^+ = \mathbb{E}\left(\sum_{i=1}^N \x_{i,k}^- \mathbb{I}_{ij}\right) = \sum_{i=1}^N \x_{i,k}^- \phi_{ij},
\end{align}
where $\phi_{ij} = p(\mathbb{I}_{ij})$, i.e., the probability that $ \x_{i,k}^-$ is mapped to $ \x_{j,k}^+$ and where the indicator $\mathbb{I}_{ij} = 1$ if $ \x_{i,k}^-$  is mapped to $ \x_{j,k}^+$  and $0$ otherwise.
$N$ denotes the number of samples being transformed.
$ \x_{i,k}^-$  indicates the $i^{th}$ sample at iteration $k$ and $ \x_{i,k}^+$  denotes the transformed sample. Compactly,
\begin{align*}
  \vo{X}_k^+ = \vo{X}_k^- \vo{\Phi} = \vo{X}_k^- N \vo{T}.
\end{align*}
$\vo{X}_k^- \in [\x_{1,k}^-, \x_{2,k}^-, \dots, \x_{N,k}^-]$ is the equally weighted  ensemble from uniform distribution $\mu$ at iteration $k$. Similarly, the equally weighted transformed ensemble from the error distribution at iteration $k$ is: $\vo{X}_k^+ \in [\x_{1,k}^+, \x_{2,k}^+, \dots, \x_{N,k}^+]$.

To ensure that the matrix $\vo{T} := [t_{ij}]$ is measure-preserving, we need to enforce the following constraints:
\begin{align}
  \sum_{i=1}^{N} t_{ij} = 1/N, \; \sum_{j=1}^N t_{ij} = w_i, \; \textrm{and} \; t_{ij} \geq 0,
\end{align}
where $\w_i$ is the equation error squared at each point $\x_{i,k}^-$.

The optimization problem can be rewritten as a linear programming problem of size $N^2$,
\begin{align}
  \vo{T}^* = \underset{T}{\arg\min} \sum_{i=1}^N \sum_{j=1}^N t_{ij} D(\x_{i,k}^-, \hat{\x}_{j,k}^+)
  \textrm{  subject to}:  \sum_{i=1}^{N} t_{ij} = 1/N, \; \sum_{j=1}^N t_{ij} = w_i, \;  t_{ij} \geq 0. \eqnlabel{ot_eqn}
\end{align}
where $ D(\x_{i,k}^-, \hat{\x}_{j,k}^+)$ is the Euclidean distance between $\x_{i,k}^-$ and $\x_{j,k}^+$.
The transformed samples $\x_{j,k}^+$ have weights equal to $w_i$.
The step, $\vo{X}_k^+ = \vo{X}_k^- \vo{\Phi} = \vo{X}_k^- N \vo{T}$ can be interpreted as a \textit{resampling}.

Our algorithm detailed in \ref{ot_alg} in section \ref{sec_otNum} makes use of this optimization to transform a uniformly weighted distribution to one that corresponds to the distribution of the square of equation error.
We then add a chosen number of samples from the transformed distribution to the training dataset and re-train our PINNs network to find the optimal solution.

\section{Numerical Simulations}
We first demonstrate the PINNs formulation to solve the steady-state FPKE with the PDE in $\eta(\x)$ as the solution variable.
In the succeeding subsection, we apply the OT-based sampling strategy to iteratively train our neural network with far fewer points than the chosen baseline and arrive at a similar degree of accuracy.

We use the same architecture for all the examples that follow: $[\textrm{d}, 48, 1]$ where $d$ is the size of the state-space of the system.
A hyperbolic tangent activation function ($\tanh$) is used between the input layer and the hidden layer, where as commonly observed in regression problems, a linear layer is used between the hidden layer and the final output layer.
We used Julia \cite{bezanson2017julia, rackauckas2017differentialequations, Flux.jl-2018, innes:2018, ma2021modelingtoolkit}
to run our simulations on the High Performance Research Computing facilities at Texas A\&M University.
To obtain the normalization constant $N_0$, we rely on numerical integration schemes.
We used the BFGS optimizer \cite{nocedal2006numerical, mogensen2018optim}  with maximum number of training iterations set to 10000 for all cases.

There are two ways to measure the accuracy of the returned solution.
For examples in which the analytical solution is available, the mean squared solution error evaluated over a finely discretized evaluation grid may be a fair criterion for judging correctness. We shall refer to this as $\epsilon_{\rho}$.
While this remains suitable for most purposes, the true test of accuracy comes from measuring the residual equation error $\epsilon_{\textrm{pde}}$ over the interior of the testing grid.
Assuming $N_{\mathcal{U}}$ is the number of points in the interior of the testing grid $\mathcal{U}$,
\begin{align}
  \epsilon_{\rho} = \frac{1}{N_{\mathcal{U}}} \sum_{i=1}^{N_{\mathcal{U}}} |{\rho}_{\Th}(\x_i) - \rho_{\textrm{true}} (\x_i)|^2, \text{ and }
  \epsilon_{\textrm{pde}} = \frac{1}{N_{\mathcal{U}}} \sum_{i=1}^{N_{\mathcal{U}}} \mathcal{N}_{\x}[\rho_{\Th}(\x_i)].
\end{align}
For systems without analytical solutions, we use $\epsilon_{\textrm{pde}}$ to gauge the accuracy of the solution.
As will be seen in the results, the stationary FPKE boundary conditions are easily satisfied.

\subsection{Steady-state FPKE using Conventional PINNs}
\label{ex_VDPR}

Consider a 2-dimensional Van der Pol - Rayleigh oscillator with dynamics given by:
\begin{align}
  \ddot{x} + (-1 + x^2 + \dot{x}^2) \dot{x} + x = \sigma \Gamma.
\end{align}
This can be rewritten as a system of differential equations in $\x = [x_1, x_2]$ as:
\begin{align*}
  \dot{x}_1 = x_2,\; \dot{x}_2 = (1 - x_1^2 - {x}_2^2){x}_2 - x_1 + \sigma \Gamma.
\end{align*}

where $\Gamma$ is standard Gaussian white noise. The FPKE in $\eta(\x)$ can be obtained from equation \ref{eqn:fkp_pde}.
The analytical solution to the stationary FPKE is available \cite{caughey1982steady}:
\begin{align}
  \rho_{ss}(\x) = N_0 \textrm{exp}\left[ \frac{1}{\sigma^2} \left((x_1^2 + x_2^2) - \frac{1}{2}(x_1^2 + x^2_2)^2 \right) \right].
\end{align}

\begin{figure}[h!]
  \centering
    \includegraphics[width =0.85 \textwidth]{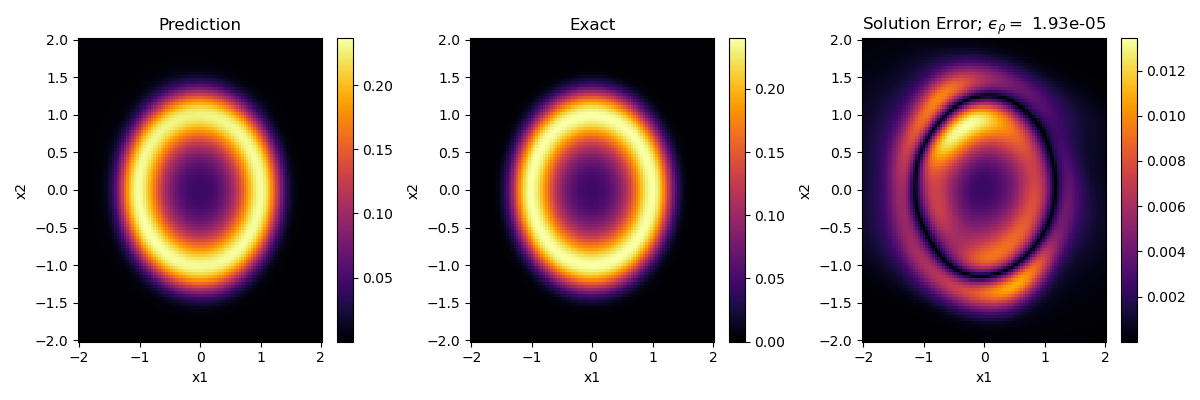}
    \caption{Van der Pol -- Rayleigh oscillator: $\rho$ for Conventional PINNs with $6241$ points.}
    \label{fig_ex2d_vdpr_pinn}
\end{figure}

We consider the domain: $\x \in [-2,2] \times [-2,2]$.
We discretize the region with $d\x = [0.05, 0.05]$. This corresponds to a training dataset of 6241 points inside the boundary of the training dataset.
After running the neural network training to get the optimal sets of weights and biases minimising the composite loss function, we obtain a solution (figure \ref{fig_ex2d_vdpr_pinn}) with  $\epsilon_{\textrm{pde}} = 1.92e^{-4}$ and $\epsilon_{\rho} = 3.5e^{-5}$.

\subsection{OT-Based Sampling for Refinement}
\label{sec_otNum}
In this section of the paper, we demonstrate the effectiveness of the proposed OT-based refinement strategy on two examples.
The first is the two-dimensional Van der Pol - Rayleigh oscillator, described above in section \ref{ex_VDPR} whose solution $\eta(\x)$ is a polynomial in state space.
The second is the classic Van der Pol oscillator, for which no analytical solution exists.

In the following algorithm, $\textrm{nOT}$ refers to the number of OT iterations we wish to carry out with $M$ points discovered in each iteration. $f(\x)$ is the dynamics of the nonlinear system in consideration without the diffusion term.
$\mathcal{S}$ and $\mathcal{U}$ denote the interiors of the training and testing set respectively.

\begin{algorithm}[h!]
\SetAlgoLined
\KwData{$\textrm{nOT}$, $f(\x)$, $\mathcal{S}$, $\mathcal{U}$, $M$, hyperparameters for NN ($\Theta$)}
  pde $\leftarrow$ Generate FPKE using \eqref{eqn:fkp_pde} for system $f(\x)$ \;
  NN $\leftarrow$ Create a PINNs using $\vo{\Theta}$ for pde \;
  $\vo{\theta}_0 \leftarrow $ Train NN with $\mathcal{S}$\;
 $i=0$\;
 \While{$i <= \textrm{nOT}$}{
  $\epsilon_{\textrm{pde}}$ $\leftarrow$ Evaluate error using equation \eqref{eqn:fkp_pde} on $\mathcal{U}$\;
  $\mathcal{U}_{\textrm{err}} \leftarrow $ Query top $M$ points corresponding to highest $\epsilon_{\textrm{pde}}$\;
  Transform $\mathcal{U}_{\textrm{err}}$ to $\mathcal{S}_{\textrm{OT}}$ using equation \eqref{eqn:ot_eqn} \;
  $\mathcal{S} \leftarrow \mathcal{S} \cup  \mathcal{S}_{\textrm{OT}}$ \;
  $i = i+1$ \;
  $\vo{\theta}_i \leftarrow $ Train NN with $\mathcal{S}$ and $\Th_{i-1}$ as the initial guess \;
 }
 \caption{OT-PINNs algorithm for solving stationary FPKE}
 \label{ot_alg}
\end{algorithm}

\subsubsection{Van der Pol - Rayleigh oscillator}

We shall assume the solution obtained above using a uniform grid of $d\x = [0.05, 0.05]$ as the baseline.
Our objective is to train a neural network with fewer points and obtain similar levels of accuracy without compromising on computational time.

For our nominal solution, we set up our network with the aforementioned architecture with $dx = 0.25$, which corresponds to 225 training points, a measly $~3.6 \%$ of the baseline training dataset.

As expected, the solution obtained after training performs poorly when its accuracy is measured w.r.t the equation error on the finely discretized testing grid.

We then solve the Optimal Transport linear programming problem to obtain 200 new data points from regions in which the solution performs the worst.
We add these points to our training dataset and subsequently train the neural network again.
Within 2 iterations of generating 200 new data points and retraining, $\epsilon_{\textrm{pde}}$ goes down to $\mathcal{O}(10^{-2})$, a remarkable increase in accuracy with just 625 data points.
Within just 5 iterations, $\epsilon_{\textrm{pde}}$ goes down to $\mathcal{O}(1e^{-4})$ with just 1225 points, i.e., less than $20 \%$ of the training dataset used for the baseline.

\begin{figure}[h!]
  \centering
  \includegraphics[width = 0.85\textwidth]{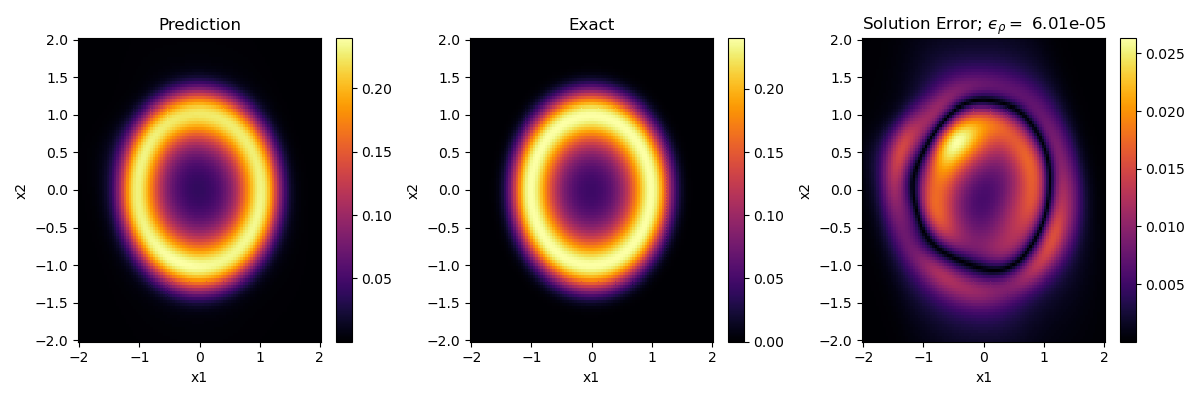}
  \caption{Van der Pol -- Rayleigh oscillator: $\rho$ for OT-PINNs with $2225$ points.}
  \label{fig_ex2d_vdpr_ot}
\end{figure}

The final solution obtained after 10 iterations of OT-based sampling can be observed in figure \ref{fig_ex2d_vdpr_ot}.
The plot of the resulting $\epsilon_{\textrm{pde}}$ after each iteration can be seen in figure \ref{EqErrOT_vdpr}.
The equation error for the nominal solution trained using 225 points is understandably large at $\mathcal{O}(1e^2)$.
After a single iteration, $\epsilon_{\textrm{pde}}$ falls to $\mathcal{O}(1e^{-1})$, a sharp decrease in magnitude from adding only 200 points.


\subsubsection{Van der Pol Oscillator}
\begin{figure}[h]
  \centering
    \includegraphics[width = \textwidth]{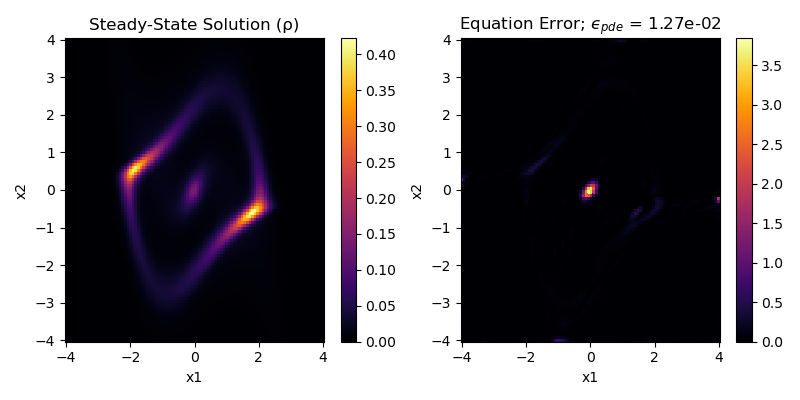}
    \caption{Van der Pol oscillator: $\rho$ for Conventional PINNs with $25281$ points.}
     \label{fig_ex2d_vdp_pinn}
\end{figure}

Consider a 2-dimensional Van der Pol oscillator with dynamics given by:
\begin{align}
  \ddot{x} + (-1 + x^2 ) \dot{x} + x = \sigma \Gamma.
\end{align}
This can be rewritten as a system of differential equations in $\x = [x_1, x_2]$ as:
\begin{align*}
  \dot{x}_1 = x_2,\;  \dot{x}_2 =  (1 - x_1^2){x}_2 - x_1 + \sigma \Gamma.
\end{align*}
We shall solve for $\rho(\x)$ with $\sigma^2 = 0.1$.
This is a more challenging example than the system described in section \ref{ex_VDPR}.
A network trained with $d\x = [0.05, 0.05]$ uses $25281$ training points inside the boundary and returns a solution with  $\epsilon_{\textrm{pde}} = 1.27e^{-2}$, as can be seen in figure \ref{fig_ex2d_vdp_pinn}.

We train our nominal network with $d\x = [0.1, 0.1]$. This corresponds to using 6241 points in a uniformly spaced grid across $[-4,4] \times [-4,4]$.
The initial $\epsilon_{\textrm{pde}}$ is tremendously high, at $\mathcal{O}(1e^{5})$.
Within 10 iterations of finding 200 new points in each step through the OT-based sampling strategy, this falls down to $\mathcal{O}(1e^{-3})$.
The proposed method has outperformed the baseline training strategy by using a fraction of the dataset and still returned a more accurate solution.

\begin{figure}[h!]
  \centering
  \includegraphics[width = \textwidth]{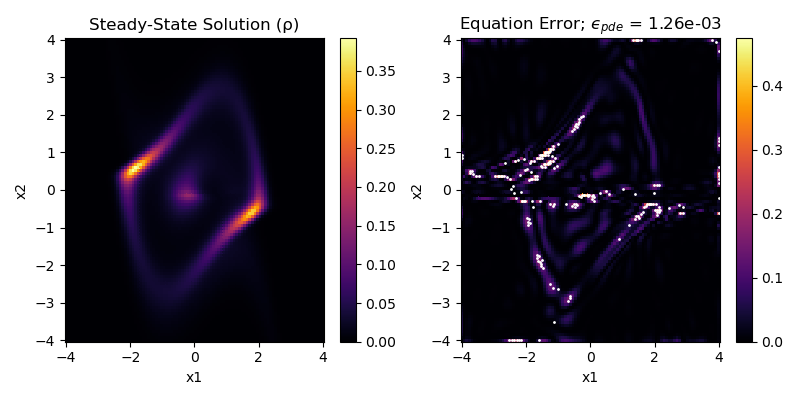}
  \caption{Van der Pol oscillator:  $\rho$ for OT-PINNs with $8241$ points.}
  \label{fig_ex2d_vdp_ot}
\end{figure}

\begin{figure}[h!]
  \centering
  \begin{subfigure}[t]{0.45\textwidth}
  \includegraphics[width = \textwidth]{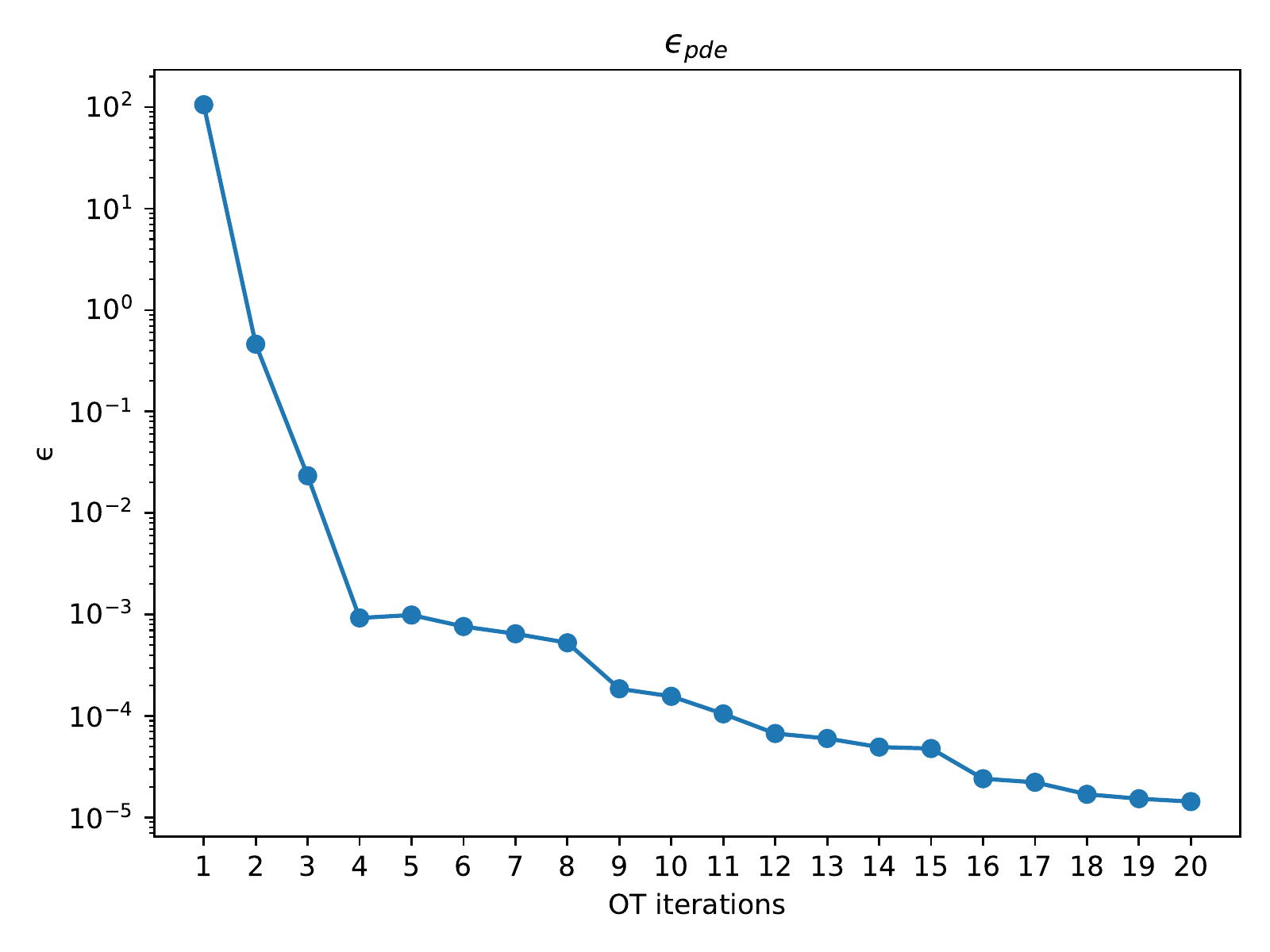}
  \caption{Van der Pol -- Rayleigh oscillator}
  \label{EqErrOT_vdpr}
  \end{subfigure}
  \centering
  \begin{subfigure}[t]{0.45\textwidth}
  \includegraphics[width = \textwidth]{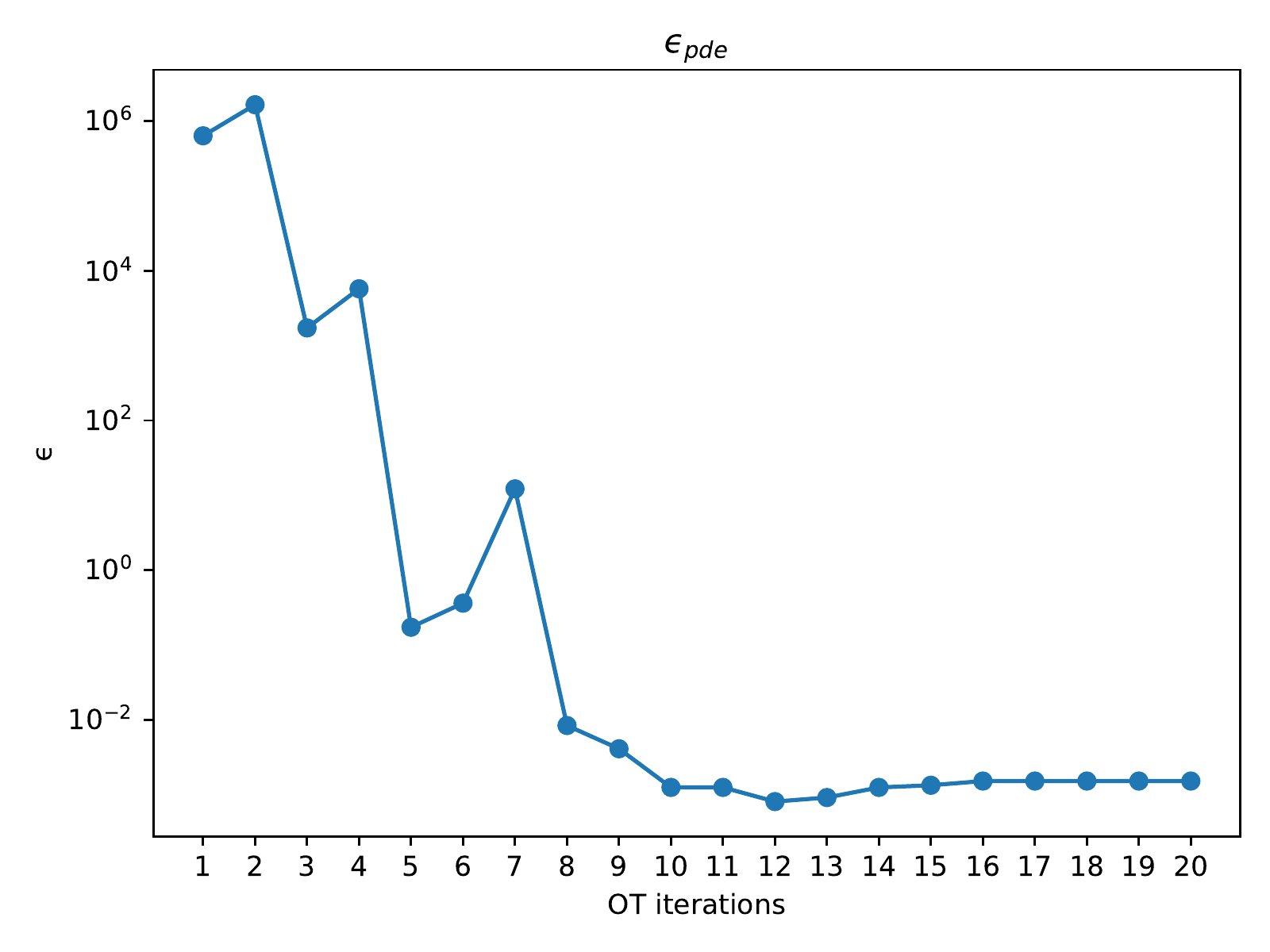}
  \caption{Van der Pol oscillator}
  \label{EqErrOT_vdp}
  \end{subfigure}
\caption{Decline of equation error with OT iterations.}
\label{EqErrOT}
\end{figure}

In figure \ref{fig_ex2d_vdp_ot} on the right, we can see the locations of the 200 points (colored white) found by solving the OT problem after 10 iterations of applying the OT technique.
The final solution can be seen on the left in figure \ref{fig_ex2d_vdp_ot}, with the limit cycle clearly visible.
The equation errors with increasing OT iterations in figure \ref{EqErrOT_vdp} follows a marginally different path than the plot for the Van der Pol - Rayleigh oscillator in figure \ref{EqErrOT_vdpr}, but the trend observed is the same.

Table \ref{results-table} summarizes the key benefits of the proposed method in terms of sample efficiency.
Clearly, the OT-PINNs algorithm is significantly more data-efficient than the baseline PINNs framework and produces comparable results.

\begin{table}[h!]
  \caption{Comparison of Training Strategies. OT-PINNs has significantly high sample-efficiency.}
  \label{results-table}
  \centering
  \begin{tabular}{p{20mm}llllc}
    \toprule
    System     & Training Method  & $N_{\mathcal{S}}$  &  $\epsilon_{\textrm{pde}}$ & $\epsilon_{\rho}$  \\
    \midrule
    Van der Pol - Rayleigh & Baseline PINNs & 6241   & $1.92e^{-4}$ & $3.5e^{-5}$  \\
         & OT-PINNs & 2225    & $1.05e^{-4}$ &  $6.01e^{-5}$ \\
    \midrule
    Van der Pol  & Baseline PINNs & 25281   & $1.27e^{-2}$ & - \\
         & OT-PINNs & 8241    & $1.26e^{-3}$ & -  \\
             \bottomrule
  \end{tabular}
\end{table}


\section{Conclusion}
In this paper, we solved the stationary Fokker-Planck-Kolmogorov equation using Physics-Informed Neural Networks.
We proposed and demonstrated an iterative strategy based on the concept of Optimal Transport to generate training datasets.
This dramatically reduced the size of dataset required to produce a solution with high accuracy.
This will pay off when we seek to study uncertainty propagation for high-dimensional systems.
Moreover, we bypassed the typical difficulties of solving the FPKE by solving an equivalent PDE in terms of the associated potential function of the probability density current for any given system.

However, there are still a few questions that remain unanswered in the proposed framework. Is there a requisite degree of accuracy that the nominal network must achieve for the OT-based strategy to succeed?
How many points should the user choose to transform and how many OT iterations would it take to return a meaningfully accurate solution?
Would increasing the number of hidden layers in the nominal network worsen the refinement strategy?
We shall investigate these questions in our future works.



\bibliographystyle{elsarticle-num}
\bibliography{citations}

\end{document}